# Electrostatic modulation of the lateral carrier density profile in field effect devices with non-linear dielectrics

Eylon Persky[1], Hyeok Yoon[2,3], Yanwu Xie[2,3,4], Harold Y. Hwang[2,3], Jonathan Ruhman[1], and Beena Kalisky[1]*.

1. Department of Physics and Institute of Nanotechnology and Advanced Materials, Bar-Ilan University, Ramat Gan 5290002, Israel.
2. Geballe Laboratory for Advanced Materials, Department of Applied Physics, Stanford University, Stanford, California 94305, USA.
3. Stanford Institute for Materials and Energy Sciences, SLAC National Accelerator Laboratory, Menlo Park, California 94025, USA.
4. Department of Physics, Zhejiang University, Hangzhou 310027, China.

*beena@biu.ac.il

We study the effects of electrostatic gating on the lateral distribution of charge carriers in two dimensional devices, in a non-linear dielectric environment. We compute the charge distribution using the Thomas-Fermi approximation to model the electrostatics of the system. The electric field lines generated by the gate are focused at the edges of the device, causing an increased depletion near the edges, compared to the center of the device. This effect strongly depends on the dimensions of the device, and the non-linear dielectric constant of the substrate. We experimentally demonstrate this effect using scanning superconducting interference device (SQUID) microscopy images of current distributions in gated $LaAlO_3/SrTiO_3$ heterostructures.

The electronic properties of two dimensional (2D) electronic systems can be effectively tuned by changing the carrier density of the system [1]. For example, a metal-insulator transition can be achieved upon removing carriers from a clean 2D system [2].

The field effect is a common way to continuously tune the carrier density of dilute 2D systems. In this approach, a metallic gate electrode is placed in proximity to the 2D system under investigation, separated by a dielectric material. The gate and the 2D system serve as two plates of a capacitor, and when a voltage is applied between them, carriers are added to, or removed from the 2D system. This approach has been utilized to study a wide variety of systems, such as semiconducting interfaces [3,4], Van der Waals heterostructures [5], and complex oxide interfaces [6].

A well-studied example of a mesoscopic system, highly tunable by the electric field effect, is the $LaAlO_3/SrTiO_3$ interface [7,8]. The large dielectric constant of the $SrTiO_3$ substrate at low temperatures allows effective tuning of the carrier density in the system; the ground state of this system can be tuned through a superconductor-insulator transition via the application of gate voltage [6]. The field effect also affects the normal transport of the system, with transitions between single- and multi- band conduction [9], tunable spin-orbit interactions [10,11], and a metal-insulator transition [12].

Quantifying the change in carrier density due to the field effect is essential in order to understand experimental data, such as magneto-transport. Further, estimating the carrier density is important to determining the relevant theoretical framework for the system. Experimentally, the charge density is often determined by measuring either the differential capacitance of the sample, or the Hall effect. However, for mesoscopic devices, the capacitive charging of the interface is inherently non-uniform, due to enhanced electric fields at the edges of the capacitor plates. Considerable attention has been paid to the vertical charge distribution, particularly in the presence of dielectric nonlinearities [13–15], and analytical calculations show that the thermodynamic properties of the 2DEG strongly depend on the vertical confinement [16,17]. However, the lateral charge distribution has not been equivalently studied. Thus, it is critical to determine the position dependent charge distribution, and quantify the spatial changes due to the finite geometry of the sample.

In this work, we used the Thomas-Fermi approximation to calculate the spatial distribution of the carriers in gated LaAlO$_3$/SrTiO$_3$ devices. We show that typical mesoscopic device geometries result in strong suppression of the carrier density at the edges of the device, suggesting that its effective width is reduced as charges are removed. We used a scanning superconducting quantum interference device (SQUID) microscope to map current paths in a gated device, and demonstrated the narrowing of the conducting region toward the center of the device, in agreement with our model. We discuss the implications of this effect to transport studies carried on the interface.

Figure 1a shows an optical picture of a typical LaAlO$_3$/SrTiO$_3$ Hall bar. Details of the epitaxial growth can be found elsewhere [18]. Devices were fabricated using standard photolithography. A long, conducting channel of width w is formed at the surface of a 5 mm wide SrTiO$_3$ substrate, with a thickness of d = 500 μm. The length of the channel is much larger than its width, so we can consider a 2D cross-section of the device (figure 1b). The gate electrode is placed at the bottom of the substrate.

To determine the electric potential, $\phi(\mathbf{r})$, we used the Laplace equation

$$\nabla[\epsilon(\mathbf{r})\nabla\phi(\mathbf{r})] = 0, \qquad (1)$$

where $\epsilon(\mathbf{r})$ is the dielectric constant. We determined the charge distribution at the interface self-consistently, using the Thomas-Fermi approximation. Within this framework, the change to the charge distribution at the interface is given by

$$\delta n(\mathbf{r}) = e\nu\big(\mu - e\phi(\mathbf{r})\big), \qquad (2)$$

where $\nu$ is the density of states, $e$ is the electron charge, and $\mu$ is the chemical potential. For a parabolic conduction band, $\nu = m^*/\pi\hbar^2$, where $m^*$ is the effective mass. The exact details of the band structure are immaterial for this calculation. For LaAlO$_3$/SrTiO$_3$, we shall assume a single, parabolic band, with $m^* = 3m_e$ [19]. We implemented equation (2) as a boundary condition, using Gauss's law:

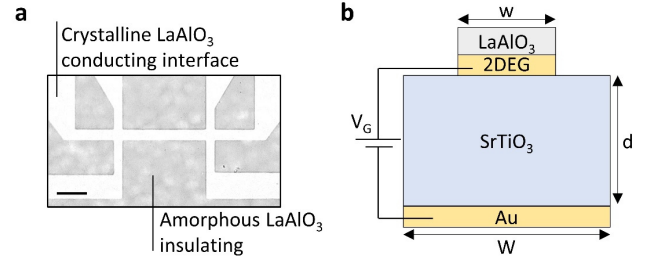

**Figure 1. (a)** Optical microscope (false color) image of a LaAlO$_3$/SrTiO$_3$ device. Bright gray areas correspond to a crystalline LaAlO$_3$ epitaxial film, beneath which a conducting interface forms, while the dark gray areas correspond to an amorphous layer, and insulating regions. Scale bar, 100 μm. **(b)** A schematic cross section (not to scale) of a patterned device. The substrate width and thickness are W and d, respectively. The width of the conducting two dimensional electron system (2DEG) is w. Gate voltage ($V_G$) is applied between the interface and a metallic electrode at the back of the substrate.

$$\epsilon\nabla\phi \cdot \hat{\mathbf{n}} = \delta n(\mathbf{r}) = \frac{m^* e^2}{\pi\hbar^2}\big(V_G + \phi(\mathbf{r})\big), \qquad (3)$$

Where $\hat{\mathbf{n}}$ is a unit vector normal to the interface plane and $V_G = \mu/e$ is the applied gate voltage. Equation (3) was obtained by considering the electric field distribution at a metallic surface. We treated the gate electrode as a perfect metal, fixing the potential at its surface to $\phi = 0$. Thus, $V_G > 0$ corresponds to removing charge from the system.

We used the multi-physics object oriented simulation environment (MOOSE) [20] to solve the Laplace equation. It is important to note that the "failed ferroelectric" state of SrTiO$_3$ at low temperatures [21,22] gives rise to a high polarizability of the substrate, resulting in a field dependent dielectric constant [23],

$$\epsilon_{SrTiO_3}(E) = 1 + \epsilon(E = 0)[1 + (E/E_0)^2]^{-1/3}, \qquad (4)$$

where, at 4 K, $\epsilon(E = 0) = 23{,}000$, and $E_0 = 82{,}000\, V/m$ [24,25]. This field dependence leads to a substantial reduction of $\epsilon$, once the gate is used: mesoscopic devices fabricated on 500 μm thick SrTiO$_3$ substrates typically require gate voltages on the order of 100 V. The resulting electric field in the substrate is ~$2\times10^5$ V/m, so that the permittivity is suppressed by a factor of ~1/2.

We first consider a w = 100 μm wide interface. Figure 2a-b show the electric field and relative permittivity profiles at the interface, resulting from a gate voltage of $V_G$ = 100 V. The electric field distribution shows two pronounced peaks, at the two edges of the conducting interface. The peaks originate from focusing of electric field lines originating from the wider (W = 5 mm) gate electrode, at the edges of the interface. To screen the enhanced field, the charge is redistributed such that a larger charge concentration is removed from the areas close to the edges, as shown in figure 2c for various gate voltages. Remarkably, the charge density at the center of this device, $\delta n(x = 0, V_G)$, can be a factor of 4 larger than the density at its edge, $\delta n(x = w/2, V_G)$.

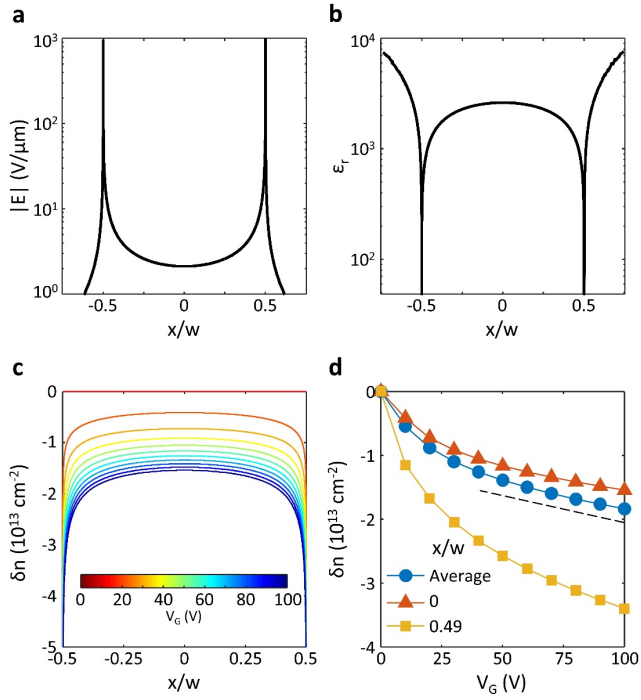
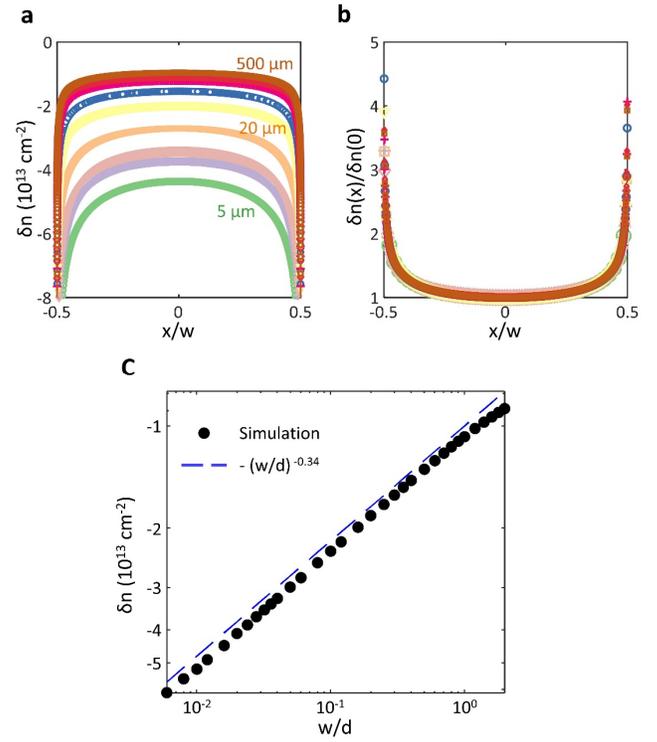

**Figure 2. (a,b)** Profiles of the electric field magnitude **(a)**, and dielectric constant **(b)**, taken along a device with w = 100 μm, at $V_G$ = 100 V. **(c)** Charge density profiles for a w = 100 μm device, for various $V_G$ values. $V_G$ changes from 0 V to 100 V, at increments of 10 V. **(d)** Changes in charge density as a function of $V_G$, at the center of the device (orange triangles), and near its edge (yellow squares). The blue circles show the device-averaged change. The black dashed line is a guide to the eye, demonstrating how the carrier density changes linearly with $V_G$, once the substrate is polarized.

**Figure 3. (a)** Charge density profiles along devices with various widths, ranging from 5 μm to 500 μm, at $V_G$ = 100 V. The substrate thickness was 500 μm. **(b)** The same profiles, rescaled by the charge density at the center of each device, showing a collapse onto a single curve. **(c)** Charge density at the center of the device, as a function of the device width, plotted on a log-log scale. The dashed line is $-(w/d)^{-0.34}$.

To understand the contribution of local variations to global measurements of the carrier density, we calculated the averaged change, $\overline{\delta n}$, for various gate voltages (figure 2c). While the charge density at the edges of the device strongly deviated from the averaged value, the density sharply increased away from the edge. Thus, the averaged value, which is readily measured in experiments, is a good measure of the carrier density at the center of the device, albeit underestimating it by ∼ 20%.

Next, we turn to study the effect of device dimensions on the spatial distribution of the charge density. We considered devices of various widths, between 5 μm and 500 μm (with w/d between 0.01 and 1). Figure 3a shows density profiles for the various geometries, for $V_G$ = 100 V. We found that, when the density is rescaled by the concentration at the center of the device, $\delta n(x = 0, V_G)$, the profiles from different devices collapse onto a single curve. Note that even for devices as large as the plate separation, the averaged density is 20% smaller than the density at the center.

While the normalized charge density is independent of the device width, the absolute change in the density varies considerably with the width. Figures 3b shows $\delta n(x = 0, V_G = 100\,V)$ as a function of the device width. For devices smaller than the plate separation, the simulations reveal a power law dependence,

$$\delta n \sim w^{-\alpha}, \quad (5)$$

with $\alpha \sim 0.34$. This nonlinear behavior is a result of the strong suppression of the dielectric constant close to the interface.

We now discuss two consequences of this analysis. First, we note that a large suppression of the carrier density at the edges of a device could be extremely important when the system is tuned through a phase transition: the edges of the sample might have a different phase than its center, because the carrier density varies across the sample.

We experimentally studied the metal-insulator transition in a $LaAlO_3/SrTiO_3$ device. We applied an alternating current of 3-30 μA RMS (frequency 1.6 kHz) to a 60 μm wide device, and imaged the resulting magnetic field using a scanning SQUID microscope, at 4 K. Figure 4a shows the magnetic flux recorded by the SQUID for various gate voltages. As we increased $V_G$, carriers were removed from the system. $LaAlO_3/SrTiO_3$ undergoes a sharp metal-insulator transition, when the carrier density is reduced below a critical threshold [12]. The magnetic flux images demonstrate that locally, the transition first occurs near the edges of the device. The current, which was initially distributed along the entire cross-section of the device, was gradually focused into the center of the device as we removed carriers from the system.

We tracked the width of the current distribution by measuring the distance between the peaks of the magnetic field distribution, as shown in figure 4b, c. The effective width of the device at $V_G = 0$ V (53 μm), was smaller than the lithographically defined width because the initial application of gate voltage (forming process) led to trapping of some of the carriers at charged impurities [15,26,27]. This effect was stronger at the edge of the device, where the electric fields generated during the forming process were stronger, leading to a reduction of the conductivity near the edges.

Assuming that the initial carrier density is $3\times10^{13}$ cm$^{-2}$, and that the threshold density for conductivity is $1.2\times10^{13}$ cm$^{-2}$ [12], we use the charge distributions calculated above to estimate the effective width of the device. The results, shown in figure 4c are in qualitative agreement with the data. The clear agreement suggests that the different critical $V_G$ required for different areas of the device are indeed attributed to the geometry-induced inhomogeneous charging. The variations between the simulation and the experimental data, at higher $V_G$, can be attributed to the presence of disorder in the sample.

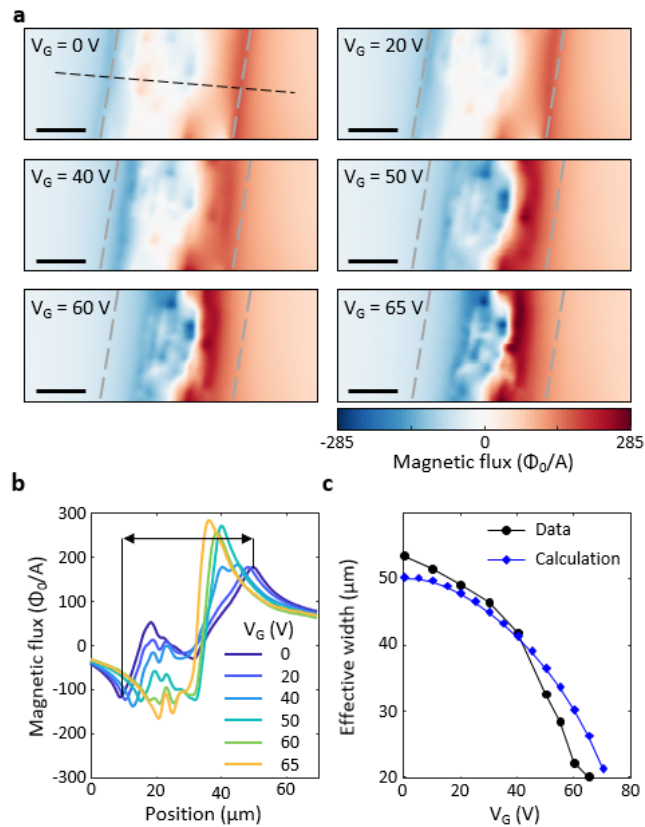

**Figure 4. (a)** Magnetic flux images of current flow in a 60 μm wide $LaAlO_3/SrTiO_3$ device, at various $V_G$. The inhomogeneity in the magnetic flux distribution inside the device is caused by disorder in the sample, and is enhanced at lower carrier densities. The gray dashed lines indicate the edges of the channel, at $V_G = 0$ V. Scale bar, 20 μm. $\Phi_0=h/2e$ is the magnetic flux quantum, and the measured flux is normalized by the applied current. **(b)** Profiles of the magnetic flux distribution at various $V_G$, taken along the black dashed line in (a). The effective width of the sample is measured as the distance between the positive and negative peaks of the magnetic field (indicated on the figure for $V_G = 0$ V). **(c)** Effective width of the device as a function of $V_G$ (black circles), compared with a calculation for a 50 μm wide homogeneous device (blue diamonds). For the simulation, we assumed the initial carrier density was $3\times10^{13}$ cm$^{-2}$, and that the threshold density required for conductivity was $1.2\times10^{13}$ cm$^{-2}$.

effectiveness of the field effects on devices with different widths. Smaller devices are more effectively tuned by the field effect than larger devices. Indeed, $LaAlO_3/SrTiO_3$ devices of various widths are commonly studied in the literature. While electronic properties of ungated structures are independent of device width [28], there are large variations in the effectiveness of the gate. Wider devices (100s of μm) typically require voltages in excess of 100 V in order to generate a change of $1\times10^{13}$ cm$^{-2}$ in the carrier density [6,9,27], while narrower devices (10s of μm) require a few dozen volts to generate the same change [12]. Devices narrower than 10 μm were shown to produce that change with a gate voltage of 2 V [29]. This survey shows that indeed, the field effect may vary dramatically in its effectiveness, due to the geometry of the device.

To conclude, we studied the effects of device geometry on the carrier distribution in mesoscopic LaAlO$_3$/SrTiO$_3$ devices. We calculated the charge density across the device, and showed that the density at the edges is dramatically suppressed, compared to the center of the device. We verified this prediction experimentally, by imaging the current distribution in a gated device, using scanning SQUID microscopy. Finally, we showed that the overall change to the charge concentration depends sub-linearly on the dimensions of the 2D system, and that this relation explains the variability reported in the literature, with regard to the performance of different field effect devices.

**Acknowledgements**

We thank A.D. Caviglia for fruitful discussions. E.P. and B.K. were supported by European Research Council Grant No. ERC-2014-STG-639792, Israeli Science Foundation grant no. ISF-1281/17, and the QuantERA ERA-NET Cofund in Quantum Technologies (Project No. 731473). H.Y., Y.X, H.Y.H were supported by the US Department of Energy, Office of Basic Energy Sciences, Division of Materials Sciences and Engineering, under contract No. DE-AC02-76SF00515.